\documentclass[11pt]{article}
\usepackage{verbatim}
\usepackage{adjustbox}
\usepackage{bbm}
\usepackage{caption}
\usepackage{epsfig}
\usepackage{graphicx}
\usepackage{subfigure}
\usepackage{bm}
\usepackage{amsbsy}
\usepackage{hyperref}
\usepackage{lscape}
\usepackage{epstopdf}
\usepackage{rotating}
\usepackage[section]{placeins}
\usepackage{amssymb}
\usepackage{lscape}
\usepackage{dsfont}
\usepackage{amsmath}
\usepackage[section]{placeins}
\usepackage{multicol}
\begin{document}
\sf
\title{Enhanced locomotion, effective diffusion, and trapping of undulatory micro-swimmers in heterogeneous environments}
\author{Arshad Kamal and Eric E Keaveny}
\maketitle
\abstract{Swimming cells and microorganisms must often move though complex fluids that contain an immersed microstructure such as polymer molecules, or filaments.  In many important biological processes, such as mammalian reproduction and bacterial infection, the size of the immersed microstructure is comparable to that of the swimming cells.  This leads to discrete swimmer-microstructure interactions that alter the swimmer's path and speed.  In this paper, we use a combination of detailed simulation and data-driven stochastic models to examine the motion of a planar undulatory swimmer in an environment of spherical obstacles tethered via linear springs to random points in the plane of locomotion.  We find that depending on environmental parameters, the interactions with the obstacles can both enhance swimming speeds, as well as prevent the swimmer from moving at all.  We also show how the discrete interactions produce translational and angular velocity fluctuations that over time lead to diffusive behaviour primarily due to the coupling of swimming and rotational diffusion.  Our results demonstrate that direct swimmer-microstructure interactions can produce changes in swimmer motion that may have important implications to the spreading of cell populations in, or the trapping of harmful pathogens by complex fluids.}

\section{Introduction}

Whether they be polymer molecules and elastic filaments, rigid and deformable particles, or even other cellular life, swimming cells and microorganisms must interact with objects immersed in the surrounding fluid and negotiate the heterogeneity that they introduce.  This situation arises in reproductive systems, such as mammalian sperm swimming through the mucin filament networks that comprise cervical mucus \cite{Rutllant2005,Ceric2005,Chretien2003}, or male gametes from the malaria parasite {\it Plasmodium} moving through dense suspensions of red blood cells in the mosquito's digestive tract \cite{Paul2002,Kuehn2010}.  This situation is also encountered in the context of disease and infection with examples including {\it H. pylori} bacteria penetrating mucus lining the stomach walls \cite{Celli2009}, or spirochetes moving through the extracellular matrix \cite{Harman2012}.  In each of these examples, the composition and density of the immersed microstructure plays a crucial role in either preventing, or allowing the cells to swim.  In the case of cervical mucus, the mucin network varies in density with the female cycle \cite{Suarez2006} and allows for the passage of the most viable sperm while trapping those with abnormal flagellar waveforms or head shapes \cite{Suarez2006,Holt2009,Holt2015}.  The filaments may even aid in guiding the sperm, helping them to navigate the reproductive tract \cite{Chretien2003}.  The trapping of small particles, cells, and viruses by mucus plays a crucial role in disease prevention, but also presents a physical barrier in drug delivery \cite{Lai2009}.

The immersed filaments or particles affect the rheological properties of the surrounding fluid, and/or create a porous environment through which the fluid must flow.  As a result, many modelling studies employ non-Newtonian constitutive laws to capture effects such as viscoelasticity \cite{Lauga2007,Fu2007,Teran2010,Gaffney2011,Spagnolie2013,Thomases2014}, shear-thinning \cite{Datt2015,Nganguia2017}, or yield stress \cite{Hewitt2017} and assess how rheology of the fluid affects swimmer motion.  The resulting changes can often be non-trivial and can depend strongly on the swimmer's stroke, as well as its ability to deform in response to stress.  For undulatory swimmers propelled by small amplitude waves, viscoelasticity hinders motion \cite{Lauga2007}, while for larger amplitudes and certain waveforms, the swimming speed can increase by a factor of about 20\% \cite{Teran2010}.  Enhanced speeds occur when the undulation period matches the relaxation time of the elastic stress \cite{Teran2010,Shen2011,Thomases2014} and further, only when the swimmer is sufficiently flexible and can be deformed by the elastic stress built up within the fluid \cite{Thomases2014}.  The effects of elasticity are be even more pronounced in gels \cite{Fu2010} where the polymer elements are not mobile as in polymer solutions, and as a result, their elastic deformation is akin to that of an elastic solid.  In this environment, the highest speeds, more than three times greater than the free swimming value \cite{Fu2010}, are obtained in the stiff limit where the governing equations reduce to those of a porous medium \cite{Leshansky2009,Leiderman2016}, for which similarly large gains in speed are observed.  

While studies using continuum models have provided key insights into how swimming speeds change with fluid rheology, they implicitly assume that the lengthscales associated with the immersed filaments, polymers or particles that produce the change in rheology are much smaller than those associated with the swimming cells.  Swimming sperm, for example, are of the same scale as the immersed filaments comprising the cervical mucus through which they swim.  The direct interactions between swimming cells and the fluid microstructure can affect swimming in different, and even more dramatic ways than those seen using continuum models.  For undulatory swimmers in networks of viscoelastic springs \cite{Wrobel2016}, hydrodynamic interactions with the network yield modest gains in speed, similar to those found with continuum models.  In fluidic environments consisting of posts arranged in a square arrays, or in wet granular media, both experiments \cite{Park2008,Juarez2010,Majmudar2012} and simulations \cite{Majmudar2012,Munch2016} demonstrate enhanced locomotion with speeds of up to ten times the free swimming value when an undulatory swimmer is able to push and pull against the posts or grains through steric interactions.  Similar results are found in simulations of helically propelled swimmers interacting with polymer elements either solely through hydrodynamic \cite{Zhang2018}, or through both hydrodynamic and steric \cite{Zoettl2017} interactions.  Along with changing the average swimming speed, direct interactions with immersed objects also introduce fluctuations, resulting in random changes in swimming speed and direction \cite{Jabbarzadeh2014,Majmudar2012,Wrobel2016}.  At long times, these fluctuations could lead to effective diffusion of the swimmers similar to that explored in the contexts of bacteria \cite{Lauga2011}, or active Brownian particles \cite{Volpe2013,Volpe2014}. 

In this paper, we explore how swimmer-microstructure interactions affect locomotion by performing numerical simulations of an undulatory micro-swimmer through a planar, random arrangement of obstacles.  Compliance is introduced by tethering each obstacle to a point in the plane via a linear spring.  This environment is intended to be a simple, planar representation of a filament network gel, with the tethers capturing network elasticity.  Our model, described in Section \ref{sec:model}, allows for hydrodynamic and steric interactions between the obstacles and swimming body.  It also accounts for swimmer deformability, thereby allowing the swimmer to change shape in response to interactions with the obstacles.  We examine in detail how obstacle density and tether strength affect swimmer motion.  Along with quantifying changes in average swimmer velocity, we also examine velocity and angular velocity fluctuations.  These results are presented in Section \ref{sec:motion}.  We then examine how these fluctuations lead to diffusive behaviour at long times.  To do this, we employ a data-driven stochastic model presented in Section \ref{sec:diffusion} to obtain expressions for the effective diffusion coefficient and correlation times and show how they change with obstacle density and tether stiffness.  Finally, in Section \ref{sec:trapping}, we examine in detail swimmer trapping, quantifying the average trapping time and how it varies with environmental parameters.  Overall, our results suggest how microstructural variations, such as those found to occur in cervical mucus during the female cycle, can allow swimming bodies to move more rapidly and diffuse through their surroundings, or stop their motion entirely.  

\section{Mathematical model for the swimmer and environment}\label{sec:model}

Our simulations are based on the mathematical model introduced in \cite{Majmudar2012} for studying undulatory locomotion through a two dimensional rigid pillar array.  The swimmer is treated as an inextensible, flexible filament of length $L$ and bending modulus $K_B$ that moves through planar undulations driven internally by a preferred curvature.  It interacts with obstacles in the plane of locomotion through hydrodynamic and steric forces.  We introduce both randomness and compliance to the environment by tethering the obstacles with linear springs to points uniformly distributed within the computational domain.  We provide a description of the model here and also refer the reader to \cite{Majmudar2012}, as well as \cite{Schoeller2018} where it was adapted to simulate sperm suspensions.

The swimmer is parametrized by arclength $s$ such that the position of a point along the swimmer is $\bm{Y}(s)$ and the unit tangent at that point is $\bm{\hat{t}} = d\bm{Y}/ds$.  Bending waves propagated along the length of the swimmer are driven by the moments per unit length, $\bm{\tau}^D = K_B \kappa_0(s,t)\bm{\hat z}$, that arise due to the preferred curvature,
\begin{equation}
\kappa_0(s,t) = K_{0} \sin\left(\frac{3\pi}{2L}s - \omega t\right) \cdot \begin{cases}
1, \hspace{20mm} s \leq L/2 \\
2(L-s) / L, \hspace{6mm} s > L/2,
\end{cases} \label{eq:PreferredCurvatureModel}
\end{equation}
where $\omega$ is the undulation frequency and $K_0$ is the amplitude.  The linear decay in the amplitude for $s > L/2$ is chosen to reproduce the waveform of the small nematode {\it C. elegans} \cite{Majmudar2012} that is often used to study locomotion in complex fluids \cite{Shen2011} and structured environments \cite{Park2008,Juarez2010,Majmudar2012}.  The swimmer is also subject to externally applied forces, $\bm{f}$, and torques, $\bm{\tau}$, per unit length that arise due to viscous stresses and steric interactions with the obstacles.  The resulting force and moment balances along the swimmer are given by
\begin{align} \label{eqn_beam1}
\frac{d \bm{\Lambda}}{d s} + \bm{f} &= 0\\ 
\frac{d \bm{M}}{d s} + \bm{\tau}^D + \bm{\hat{t}}\times \bm{\Lambda}+\bm{\tau} &= 0. \label{eqn_beam2}
\end{align}
where $\bm{\Lambda}$ is the internal stress that enforces inextensibility and $\bm{M} = K_B \bm{\hat{t}}\times d\bm{\hat{t}}/ds$ is the bending moment.  

To obtain a numerical solution to these equations, the swimmer is discretised into $N$ segments of length $\Delta L = L/N$ with the position of segment $n$ given by $\bm{Y}_{n}$, while the tangent at that point is denoted as $\bm{\hat {t}}_{n}$.  Taking $\bm{\Lambda}$ and $\bm{M}$ at the midpoints between adjacent segments, and replacing the differential operator in Eqs. (\ref{eqn_beam1}) and (\ref{eqn_beam2}) by central finite differences, we obtain the following discretised system
\begin{align}
\frac{\bm{\Lambda}_{n+1/2} - \bm{\Lambda}_{n-1/2}}{\Delta L} + \bm{f}_n &= 0 \label{eq:disfbal} \\
\frac{\bm{M}_{n+1/2} - \bm{M}_{n-1/2}}{\Delta L} +
\frac{1}{2}\bm{\hat{t}}_n\times (\bm{\Lambda}_{n+1/2} + \bm{\Lambda}_{n-1/2})+\bm{\tau}^D_n + \bm{\tau}_n &= 0 \label{eqn:dismbal},
\end{align}
where $\bm{M}_{n+1/2} = (K_B/\Delta L) \bm{\hat{t}}_n \times \bm{\hat{t}}_{n+1}$.  For this discrete system, $\bm{\Lambda}_{n+1/2}$ is the Lagrange multiplier that enforces the discrete version of the inextensibility constraints,
\begin{align}
\bm{Y}_{n+1} - \bm{Y}_{n} - \frac{\Delta L}{2} (\bm{\hat{t}}_{n+1} + \bm{\hat{t}}_{n}) = \bm{0}. \label{eq:cons}
\end{align}
Multiplying Eqs. (\ref{eq:disfbal}) and (\ref{eqn:dismbal}) through by $\Delta L$, we obtain the force and moment balances for each of the segments.  For segment $n$, we have
\begin{align}
	\bm{F}^{C}_n  + \bm{F}^{H}_n + \bm{F}^{S}_{n} &= 0, \label{eqn_force_balance}\\
	\bm{T}^{B}_n + \bm{T}^{C}_n  +  \bm{T}^{D}_n + \bm{T}^{H}_n &= 0.\label{eqn_torque_balance}
\end{align}
where $\bm{F}^{C}_n =  \bm{\Lambda}_{n+1/2} - \bm{\Lambda}_{n-1/2}$, $\bm{T}^{B}_n =  \bm{M}_{n+1/2} - \bm{M}_{n-1/2}$, and $\bm{T}^{C}_n = (\Delta L/2)\bm{\hat{t}}_n\times (\bm{\Lambda}_{n+1/2} + \bm{\Lambda}_{n-1/2})$.  The hydrodynamic forces, $\bm{F}^{H}_n$, and those due to steric interactions with the obstacles, $\bm{F}_{n}^{S}$, are the total external force on the segment $n$ such that $\bm{F}^{H}_n+\bm{F}_{n}^{S} = \bm{f}_n\Delta L$, and the hydrodynamic torques $ \bm{T}^{H}_n = \bm{\tau}_n\Delta L$ are the only external torques on the segments.  Finally, $\bm{T}^{D}_n = \bm{\tau}^D_n \Delta L$ are the torques due to the preferred curvature $\kappa_0$,  Eq. (\ref{eq:PreferredCurvatureModel}), and are given by
$\bm{T}^D_n= K_{B}(\kappa(s_n,t) - \kappa(s_{n+1},t)) \hat{\bm{z}}$, where $s_n = (n - 1/2)\Delta L$.

Each obstacle, $n$, is a sphere of radius $A$ tethered to a point $\bm{X}_n$ by a linear spring, such that the tether force is 
\begin{equation}
{\bm F}^{T}_n = - k\left({\bm Y}_n - {\bm X}_{n}\right), \label{eq:SpringForce}
\end{equation}
where ${\bm Y}_n$ is the position of the obstacle and $k$ is the spring constant.  In addition to this tether force, each obstacle will experience hydrodynamic forces, ${\bm F}_{n}^{H}$, due to the surrounding fluid, as well as steric forces, ${\bm F}_{n}^{S}$, with the swimmer and/or other obstacles.  The resulting force balance for obstacle $n$ is then
\begin{equation}
{\bm F}_{n}^{H} + {\bm F}_{n}^{T} + {\bm F}_{n}^{S} = {\bm 0}. \label{eq:ObstaclesForceBalance}
\end{equation}
The obstacles are taken to be torque-free.  

The obstacles and swimmer segments interact with each other through the steric and hydrodynamics forces that appear in their respective force and torque balances.  The steric forces between obstacles and swimmer segments, as well as those between obstacles, are captured through a short-ranged, pair-wise repulsive barrier force \cite{Dance2004}.  The force on obstacle or segment $n$ due to obstacle or segment $m$ is 
\begin{align*}
	\bm{F}^S_{nm} &=F_{nm} \left(\frac{(\chi R_{nm})^2 - \left|\bm{Y}_n -\bm{Y}_{m}\right|^2}{(\chi R_{nm})^2 - R_{nm}^2}\right)^4 \frac{\left(\bm{Y}_n - \bm{Y}_{m}\right)}{2R_{nm}},
\end{align*}
if $\left|\bm{Y}_n -\bm{Y}_{m}\right| < \chi R_{nm}$, and zero otherwise.  The parameter $F_{nm}$ sets the strength of the force at contact and $\chi$ controls the range over which force acts.  $R_{nm}$ in the distance between $n$ and $m$ at contact.   For obstacle-obstacle interactions, $R_{nm} = 2A$ and $F_{nm} = 152 K_B/L^2$, while for segment-obstacle interactions, $R_{nm} = a + A$ and $F_{nm} = 57 K_B/L^2$ with the segment radius being $a = \Delta L/2.2$.  For all interactions, we take $\chi = 1.1$.  

Hydrodynamic interactions are incorporated by considering the coupled low Reynolds number mobility problem established by the force and moment balances for the segments, Eqs. (\ref{eqn_force_balance}) and (\ref{eqn_torque_balance}), together with that for the obstacles, Eq. (\ref{eq:ObstaclesForceBalance}).  We employ the force-coupling method (FCM) \cite{Maxey2001, Lomholt2003, Liu2009} to solve the mobility problem and obtain the translational and angular motion of the segments and obstacles.  In FCM, the forces and torques the segments and obstacles exert on the fluid are treated though a low-order finite-force multipole expansion in the Stokes equations and the resulting fluid flow due to this forcing is volume averaged to obtain the velocity, ${\bm U}_{n}$, and angular velocity, ${\bm \Omega}_{n}$, for each particle $n$.  

After obtaining the motion of the obstacles and segments, we update their positions and orientations.  As swimmer deformation is restricted to a plane, we know $\bm{\Omega}_n = \Omega_n \bm{\hat z}$ and may introduce an angle $\theta_n$ for each segment $n$, such that $\bm{\hat t}_n = (\cos \theta_n, \sin \theta_n)$.  Therefore, to update particle positions and orientations, we integrate in time
\begin{align}
\frac{d\bm{Y}_n}{dt} = \bm{U}_n \\
\frac{d \theta_n}{d t} = \Omega_n,  
\end{align}
while simultaneously obtaining the Lagrange multipliers to ensure the inextensibility constraints, Eq. (\ref{eq:cons}), are satisfied.  To do this, we employ the second-order implicit backward differentiation scheme \cite{Ascher1998} to integrate the differential equations, and Broyden's method \cite{Broyden1965} to solve the resulting system of equations for the updated values of $\bm{Y}_n$, $\theta_n$, and the Lagrange multipliers.  

\begin{figure}[h]
\centering
\includegraphics[trim={0 0 0 5cm}, scale=0.15]{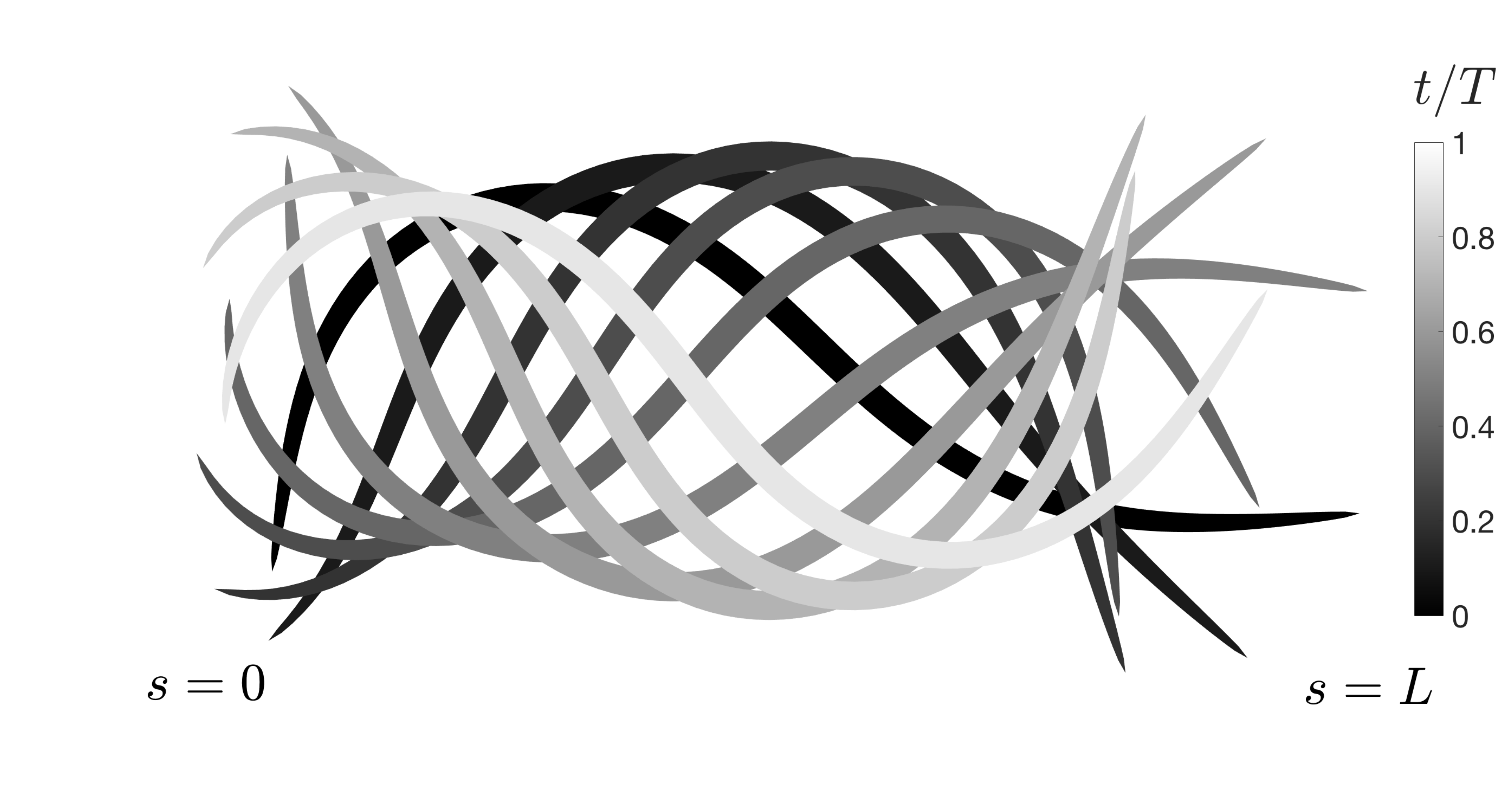} %EPS file
\captionsetup{width=0.9\textwidth}
\caption{\sf Swimmer shape over for one period of undulation.  The swimmer is moving to the left and the gray level fades as time progresses.}
\label{fig:WaveCharacteristicsImageFigure}
\end{figure}

In our simulations, as in \cite{Majmudar2012}, the swimmer is discretised into $N = 15$ segments and the preferred curvature amplitude is $K_{0} = 8.25/L$.  The frequency, $\omega$, of the preferred curvature wave sets the dimensionless sperm number to be, $Sp = (4\pi \omega \eta / K_{B})^{1/4} L \approx 5.87$, where $\eta$ is the viscosity of the fluid.  The sperm number \cite{Lowe2003, Lauga2009} provides a measure of the ratio of the viscous and elastic forces acting on the swimmer.  The corresponding waveform for the swimmer over a single undulation period, $T = 2\pi / \omega$, is shown in Fig. \ref{fig:WaveCharacteristicsImageFigure}.  The swimming speed in the absence of obstacles is found to be $U_{0} = 0.01225 \omega L$.  

The simulations presented in the subsequent sections are performed in periodic domains of size $L_D \times L_D \times L_z$, where we have $L_D = 2.53 L$ for short-time simulations described in Sections \ref{sec:motion}, and $L_D = 7.06 L$ for our longer-time simulations shown in Section \ref{sec:diffusion}.  In both cases, the out of plane thickness of the domain is $L_z = 0.29L$.  The obstacle radius is set to $A = 0.061L$.  To vary the stiffness of the environment, we adjust the non-dimensional parameter
\begin{align}
k_{sp} = kL^3/K_{B}
\end{align}
which describes the strength of the tether spring constant relative to swimmer stiffness.   The obstacle density is controlled by the in-plane area fraction,
\begin{align}
\varphi = N_{obs}\pi A^2/L_D^2,
\end{align}
where $N_{obs}$ is the number of obstacles.

\section{Locomotion speed and induced velocity fluctuations vary with the obstacle density and stiffness}\label{sec:motion}

\begin{figure}[h]
\centering
\includegraphics[scale=0.32]{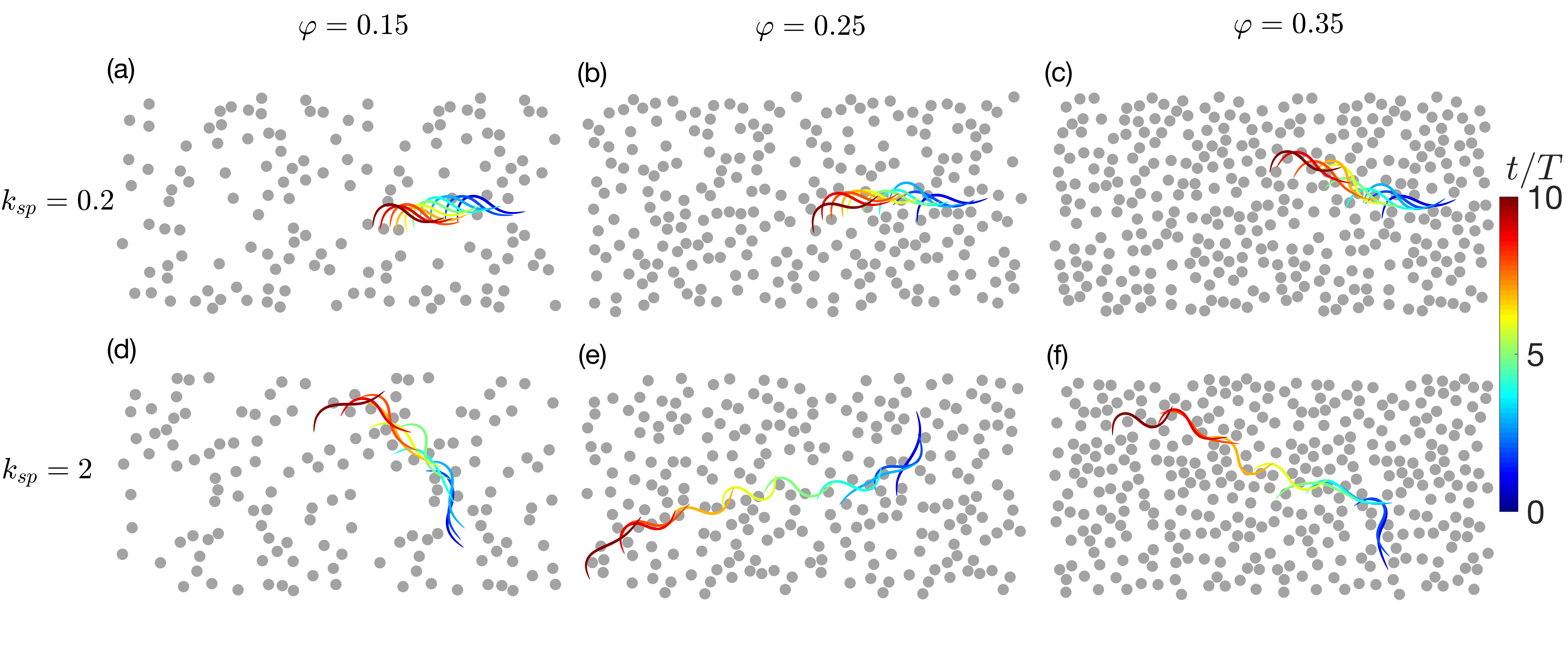}  %EPS file
\captionsetup{width=0.9\textwidth}
\caption{\sf Swimmer motion for $10$ undulation periods in environments with obstacle densities $\varphi = 0.15, 0.25$ $\&$ $0.35$ and tether stiffnesses $k_{sp} = 0.2$ (a - c) and $k_{sp} = 2$ (d - f). The figures show the obstacles at their tether points.}
\label{fig:ShortTimeTrajectoriesPlotFigure}
\end{figure}

We begin by presenting results from short-time simulations performed for a range of obstacle densities and tether stiffnesses.  Each simulation is run for ten undulation periods, over which time swimmer motion is recorded and analysed.  Fig. \ref{fig:ShortTimeTrajectoriesPlotFigure} shows the obstacle rest configuration and the swimmer after each period from representative simulations with $k_{sp} = 0.2$ and $2$, and for $\varphi = 0.15, 0.25$, and $0.35$.  Videos of the swimmer moving through different environments are included in the electronic Supplementary Materials.  We observe that when the medium is relatively compliant $(k_{sp} = 0.2)$ and the obstacle density is low $(\varphi = 0.15)$, the swimmer moves in a straight line and its shape is nearly identical after each period.  When the obstacle density is increased to $\varphi = 0.25$, the swimmer moves, on average, in a line, but now covers more distance per period, and there are noticeable fluctuations in the swimmer position from period to period.  These changes become more pronounced when the density is increased to $\varphi = 0.35$.  In the less compliant environment ($k_{sp} = 2$), we see that even for low obstacle densities, the swimming direction is affected by the presence of the obstacles.  We also observe now that the swimmer shape varies from period-to-period due to interactions with the obstacles, and at higher obstacle densities, the swimmer moves significantly greater distances than in the more compliant environment, approaching one swimmer length in one undulation period.  

To quantify effects of obstacle density and tether stiffness on swimmer motion, we examine the means and covariances of the swimmer's period-averaged velocity and angular velocity obtained from $40$ independent, short-time simulations for different $k_{sp}$ and $\varphi$.  By examining period-averaged quantities, we eliminate artificial contributions to the covariances due to periodic variations in the swimmer's velocity and angular velocity as a result of its periodic change in shape.   Specifically, at each time $t$, we determine the swimmer's instantaneous centre-of-mass velocity
\begin{align}
{\bm V} = \frac{1}{N} \sum_{n=1}^{N} {\bm U}_{n}.
\end{align}
and instantaneous orientation, ${\hat {\bm q}} = {\bm q}/q$, where
\begin{align}
{\bm q} = -\frac{1}{N} \sum_{n=1}^{N} {\hat {\bm t}}_{n}. \label{eq:qVectorDefinition}
\end{align}
and $q = \lvert {\bm q}\rvert$.  Defining the swimmer's instantaneous angular velocity through $d\bm{\hat q}/dt = \Omega \bm{\hat z} \times \bm{\hat q}$, we obtain the following relation between $\Omega$ and the angular velocity of each segment,
\begin{equation}
\Omega = -\frac{1}{Nq} \left( \sum_{n=1}^{N_{w}} \Omega_{n} ( {\hat {\bm q}} \cdot {\hat {\bm t}}_{n}) \right). \label{eq:CofMAngVelDefinition}
\end{equation}

From these instantaneous values, we determine their period-averaged counterparts, which for period $i$ are given by 
\begin{align}
{\bm V}_{i} = \frac{1}{T} \int_{(i-1)T}^{iT} {\bm V}(t) dt, \label{eq:PeriodAveragedVelocity} \\
\Omega_{i} = \frac{1}{T} \int_{(i-1)T}^{iT} \Omega(t) dt,\label{eq:PeriodAveragedAngVelocity} \\
{\bm q}_{i} = \frac{1}{T} \int_{(i-1)T}^{iT} {\hat {\bm q}}(t) dt, \label{eq:PeriodAveragedOrientation1} 
\end{align}
with the period-averaged swimmer orientation being ${\hat {\bm p}}_{i} = {\bm q}_{i} /\lvert {\bm q}_{i} \rvert$.  From these quantities, we obtain the swimmer velocity in the body frame $V_{p,i} = \bm{V}_i \cdot \bm{\hat p}_i$ and $V_{n,i} = \bm{V}_i \cdot \bm{\hat n}_i$, where $\bm{\hat n}_i =  \bm{\hat z} \times \bm{\hat p}_i$.  We then compute their averages, $ \langle V_{p} \rangle$ and $ \langle V_{n} \rangle$, respectively, as well as the average angular velocity, $\langle \Omega \rangle$ and the $3\times 3$ covariance matrix 
\begin{align}\label{eq:covarmat}
\bm{C} = \langle \bm{W}\bm{W}^T \rangle -  \langle\bm{W}\rangle\langle\bm{W}^T \rangle,
\end{align} 
where $\bm{W} = (V_p, V_n, \Omega)^T$.  In these expressions, the angular brackets, $\langle \cdot \rangle$, denotes the expectation, which in our case is computed by averaging the quantity over the final $8$ undulation periods of the $40$ independent simulations for each value of $k_{sp}$ and $\varphi$.  

\subsection{Locomotion speed}

\begin{figure}[h]
\centering
 \includegraphics[width=5in]{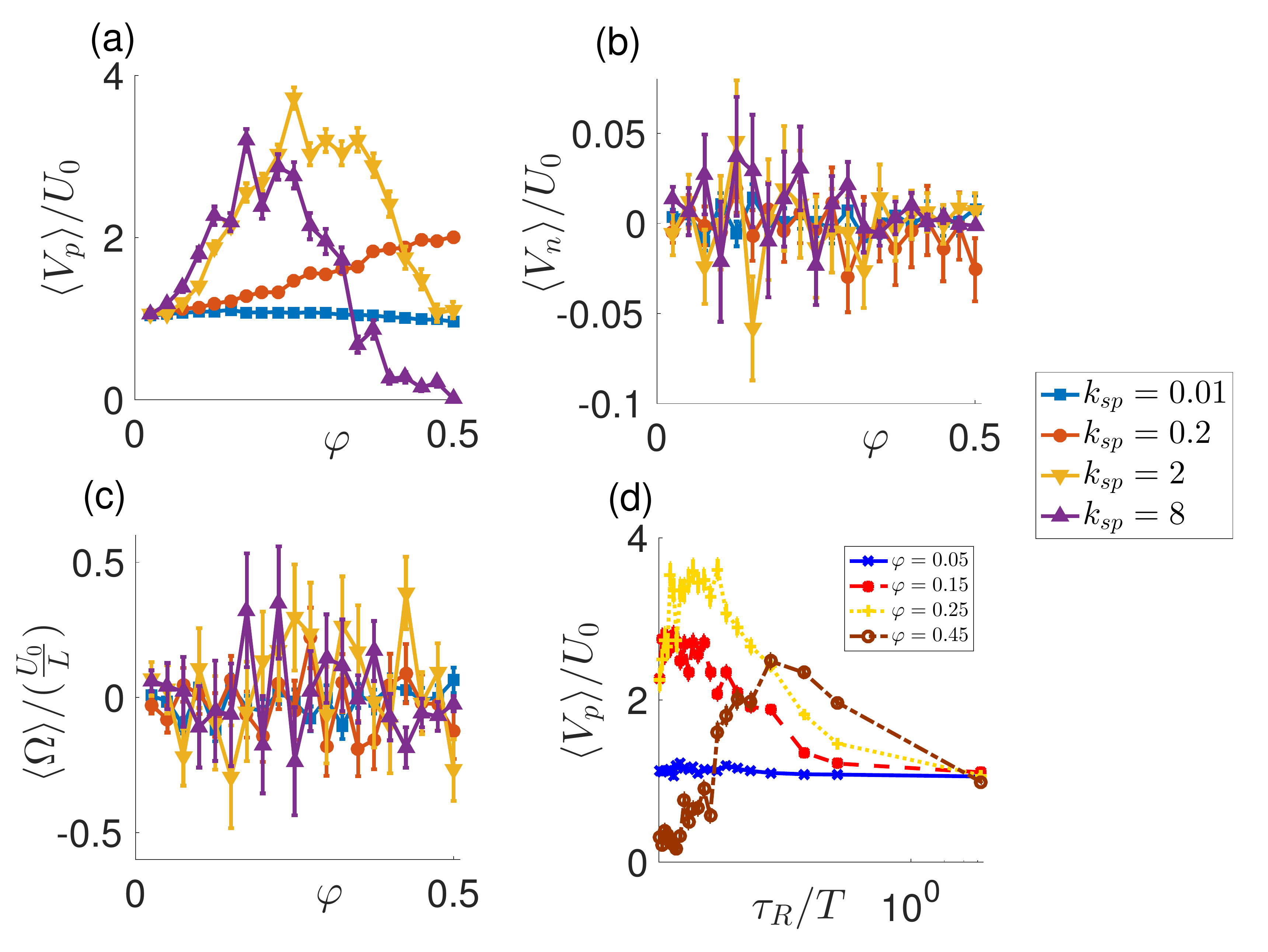}
 \captionsetup{width=0.9\textwidth}
 \caption{\sf Average swimming speed, (a) $\langle V_p\rangle$, normal velocity, (b) $\langle V_n\rangle$, and angular velocity, (c) $\langle \Omega \rangle$, versus $\varphi$ for tether stiffness $k_{sp} = 0.01, 0.2, 2$ $\&$ $8$.  Panel (d) shows $\langle V_p\rangle$ as a function of the tether relaxation time, $\tau_R$, for different $\varphi$.}
\label{fig:Averages}
\end{figure}

Fig. \ref{fig:Averages} shows $\langle V_p\rangle$, $\langle V_n\rangle$, and $\langle \Omega \rangle$ for tether stiffnesses $k_{sp} = 0.01, 0.2, 2,$ and $8$ and for obstacle densities ranging from $\varphi = 0.025$ to $\varphi = 0.5$.   We see that for all $\varphi$ and $k_{sp}$, the swimmer moves, on average, in the direction $\bm{\hat p}$ and there is no average swimmer rotation.  When the tether stiffness is very low, $k_{sp} = 0.01$, we find that the motion is slightly hindered by the presence of the obstacles, with the speed decreasing monotonically with obstacle density to a value of $\langle V_{p} \rangle = 0.965 U_{0}$ at $\varphi = 0.5$.  Increasing the stiffness to $k_{sp} = 0.2$, we now observe that swimming is enhanced by the obstacles.  The speed increases linearly with obstacle density and reaches a value of nearly double its free-swimming speed at $\varphi = 0.5$.  

For tether stiffnesses $k_{sp} = 2$ and $k_{sp} = 8$, the swimming speed can reach even larger values, as well as exhibit a more complex, non-monotonic dependence on $\varphi$.  The maximum swimming speeds we observe are $\langle V_{p} \rangle = 3.72 U_{0}$ for $k_{sp} = 2$, and $\langle V_{p} \rangle = 3.2 U_{0}$ for $k_{sp} = 8$ and occur at $\varphi = 0.25$ and $\varphi = 0.175$, respectively.  These values are much larger than the modest increases of 20\% observed for undulatory swimming in continuous viscoelastic fluids \cite{Teran2010,Shen2011} and viscoelastic networks \cite{Wrobel2016}, though very close to the enhanced speeds found using continuum descriptions of gel networks \cite{Fu2010} and in Brinkman fluids \cite{Leshansky2009,Leiderman2016}.  Our results are also consistent with the trends found with these continuum models for which stiffer environments lead to faster speeds, especially when the swimmer shape changes in response to the environment \cite{Thomases2014}.  As in structured environments \cite{Park2008,Majmudar2012}, the mechanism behind the increase in speed is that the swimming body is able to push and pull against the obstacles to overcome the force-free constraint imposed by low Reynolds number swimming.  

At high obstacle densities, we observe a reduction in speed for these stiffer systems.  We note that this is not due to a uniform reduction across all independent simulations, but rather the result of the swimmer becoming completely trapped by the environment in a subset of the simulations.  In the most extreme case where $k_{sp} = 8$ and $\varphi = 0.5$, nearly all swimmers are trapped instantaneously and the average speed is very close to zero.  We have also performed averaging with the trapped cases excluded (see Supplementary Material), and though we do still observe a decrease in the swimming speed at large $\varphi$, only for $k_{sp} = 8$ and $\varphi = 0.5$ do we find that the speed is less than the free swimming value with $\langle V_{p} \rangle = 0.51U_{0}$.  We note, however, that this value arises from a single simulation, and even in that case, the swimmer became trapped after two periods of measurement. 

In addition to measuring tether stiffness relative to that of the swimming body through $k_{sp}$, we may instead examine how the swimming speed varies with the obstacle relaxation time, $\tau_R =  6\pi A\eta/k$, given by the ratio of the obstacle drag coefficient to the tether spring constant.  Fig. \ref{fig:Averages}d shows the swimming speed as a function of $\tau_R$ for obstacle densities ranging from $\varphi = 0.05$ to $0.45$.  For low obstacle densities, we see only modest increases in swimming speeds as the environment becomes stiffer ($\tau_R \rightarrow 0$).  As $\varphi$ increases, the enhancement in the swimming speed becomes more dramatic, which for $\varphi = 0.25$, closely resembles the dependence on $\tau_R$ obtained for swimming sheets in continuum models of gels \cite{Fu2010}.  At the highest obstacle density, $\varphi = 0.45$, we observe enhanced speeds when the relaxation time is large, with the highest value $\langle V_{p} \rangle = 2.5U_{0}$ occurring $\tau_R/T = 0.054$.  Decreasing $\tau_R$ below this value, the speed drops substantially due swimmer trapping.  

\subsection{Velocity fluctuations}

\begin{figure}[h]
\centering
  \includegraphics[width=7in]{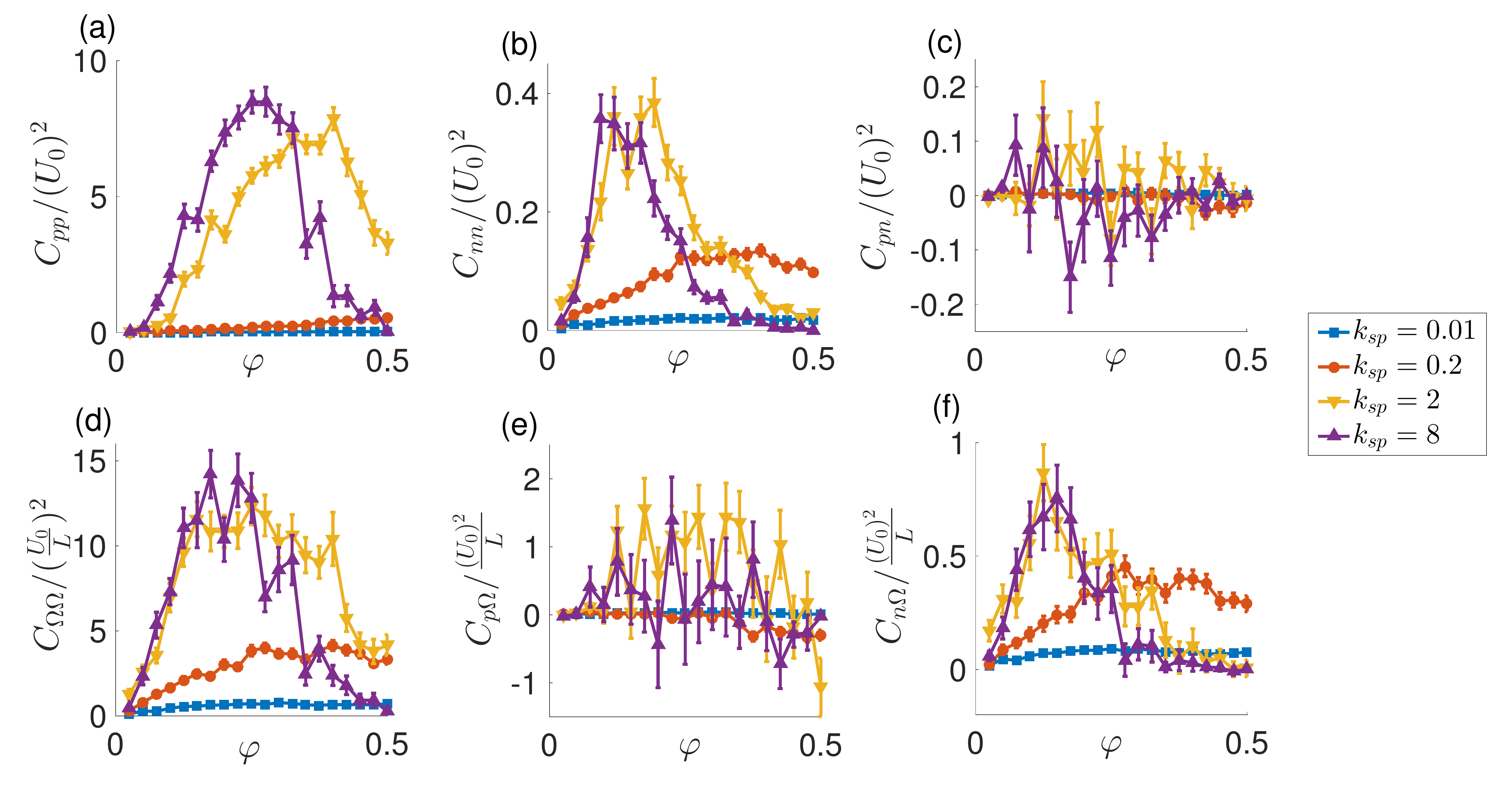}
  \captionsetup{width=0.9\textwidth}
\caption{\sf Entries of the velocity covariance matrix, $\bm{C}$, versus $\varphi$ for $k_{sp} = 0.01, 0.2, 2$ $\&$ $8$. Panels (a - c) show the translational velocity entries (a) $C_{pp}$, (b) $C_{nn}$, and (c) $C_{pn}$, while panels (d -- f) show the angular velocity covariance (d) $C_{\Omega\Omega}$ and the translational-rotational velocity covariances (e) $C_{p\Omega}$ and (f) $C_{n\Omega}$.}
\label{fig:Fluctuations}
\end{figure}

Along with changes in the swimmer's average motion, the discrete interactions with the obstacles lead to fluctuations in the translational and angular velocities.  Fig. \ref{fig:Fluctuations} shows the entries of the covariance matrix, $\bm{C}$ as a function of $\varphi$ for $k_{sp} = 0.01, 0.2, 2,$ and $8$.  We find that in the body frame, the translational-translational velocity covariance is diagonal as the entry $C_{np}$ is nearly zero for each value of $k_{sp}$ across the entire range of $\varphi$.  We see, however, that the velocity fluctuations are anisotropic as the maximum values of $C_{pp}$ are more than an order of magnitude greater than those of $C_{nn}$.  As with the swimming speed, when $k_{sp} = 0.01$, the presence of the obstacles has little effect on swimmer motion and the entries of $\bm{C}$ remain very close to zero.  For $k_{sp} = 0.2$, the entries $C_{pp}$ and $C_{nn}$ grow with $\varphi$, though for $C_{nn}$, this growth stops at approximately $\varphi = 0.3$ and $C_{nn}$ remains constant at higher $\varphi$.  When the tether stiffness is high ($k_{sp} = 2$ and $k_{sp}=8$), $C_{pp}$ and $C_{nn}$ exhibit a non-monotonic dependence on $\varphi$ due to swimmer trapping.

In addition to translational motion, we find significant angular velocity fluctuations due to interactions with the obstacles.  The values of $C_{\Omega\Omega}$ are comparable in magnitude to the translational velocity fluctuations and exhibit a similar dependence with $\varphi$ as $C_{pp}$ and $C_{nn}$.  Interestingly, we also find that the off-diagonal entry, $C_{n\Omega}$ and, to a lesser extent $C_{p\Omega}$, which provide the covariance of the swimmer's translational and rotational motion is, in general, non-zero.  This indicates that when the swimmer pushed by the obstacles in the $\bm{\hat{n}}$-direction, it also tends to be rotated anticlockwise. 

\section{Diffusive behaviour at long times is characterised by rotational diffusion and forward locomotion}
\label{sec:diffusion}
\begin{figure}[h]
\centering
 \includegraphics[width=5in]{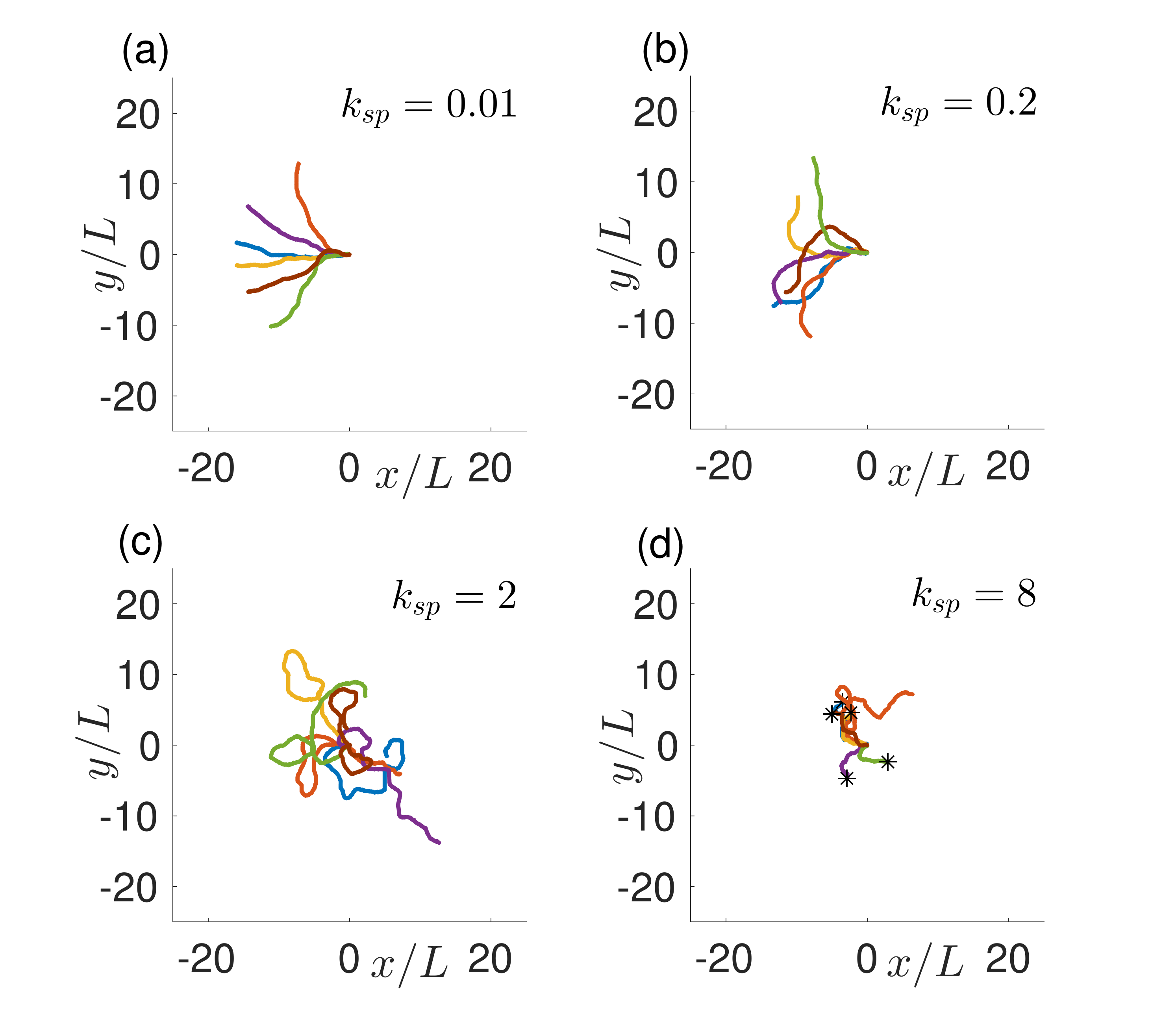}
 \captionsetup{width=0.9\textwidth}
\caption{\sf Swimmer trajectories for $200$ undulation periods for (a) $k_{sp} = 0.01$, (b) $0.2$, (c) $2$ and (d) $8$ with obstacle density $\varphi = 0.15$. Six independent paths are displayed for each case.  Asterisks in panel (d) show the location where the swimmer was trapped by the obstacles.}
\label{fig:TrajectoriesKsp}
\end{figure}

\begin{figure}[h]
\centering
 \includegraphics[width=5in]{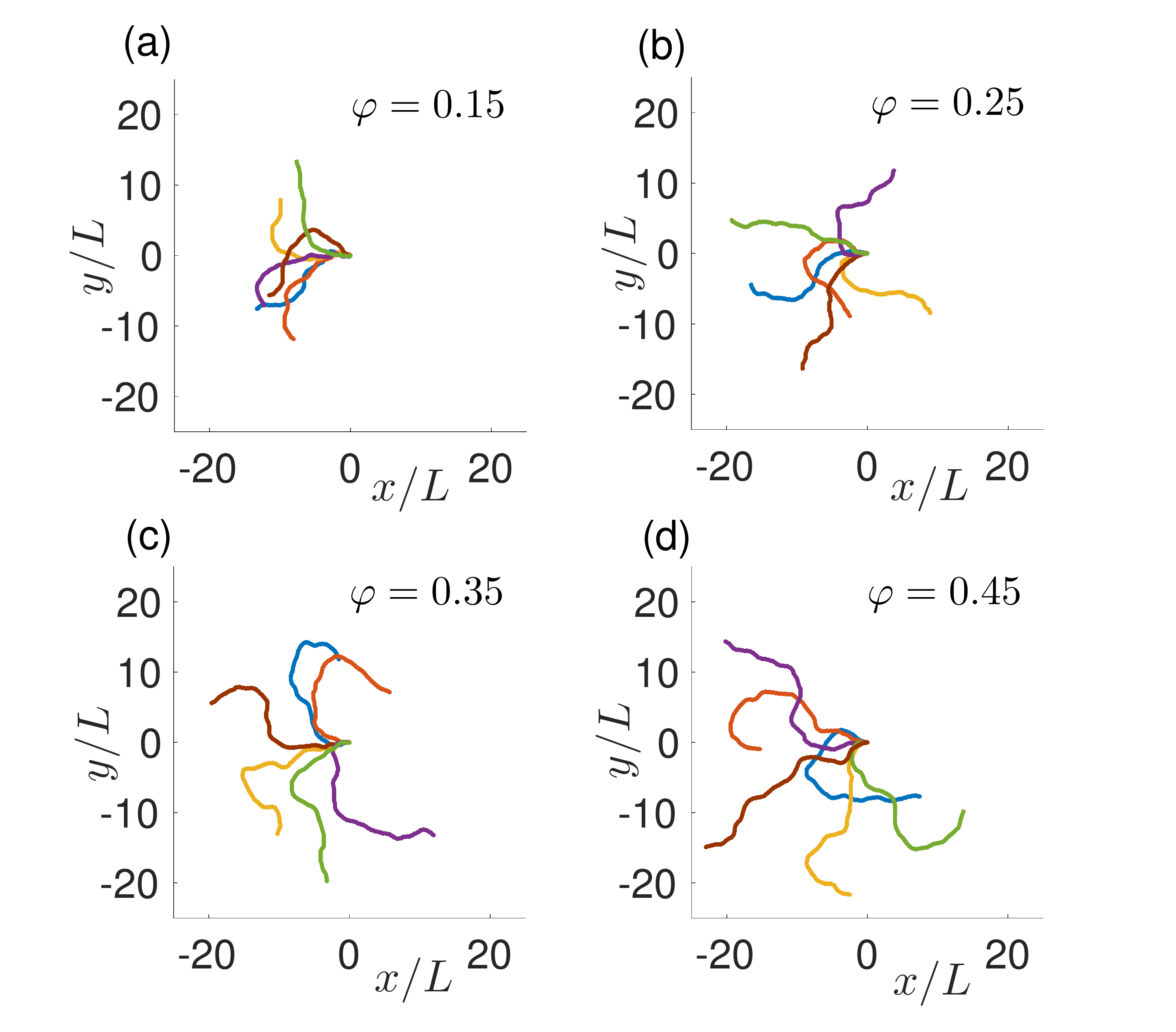}
 \captionsetup{width=0.9\textwidth}
\caption{\sf Swimmer trajectories for $200$ undulation periods for (a) $\varphi = 0.15$, (b) $0.25$, (c) $0.35$, and (d) $0.45$ with obstacle density $k_{sp} = 0.2$. Six independent paths are displayed for each case. }
\label{fig:TrajectoriesAf}
\end{figure}

The translational and angular velocity fluctuations due to collisions with the obstacles presented in the previous section can, over longer times, result in the swimmer exhibiting a random walk that can be characterized by an effective diffusion coefficient.  This is a distinct difference from motion through continuous environments, even non-Newtonian ones, for which, in the absence of boundaries, a swimming body undergoing symmetric, periodic shape changes moves in a straight path.  Trajectories from simulations run for $200$ undulation periods with $\varphi = 0.15$ and $k_{sp} = 0.01, 0.2, 2$, and $8$ are shown in Fig. \ref{fig:TrajectoriesKsp}.  In each plot and for each trajectory, the swimmer's centre-of-mass is initially located at the origin and swimming to the left.  To avoid the swimmer exhibiting periodic, though complicated, trajectories, when performing these simulations we intermittently reseed obstacles far away from the swimmer using the procedure described in the Supplementary Materials.  From Fig. \ref{fig:TrajectoriesKsp}, we see that increasing the stiffness of the tethers results in more frequent and sharper turns, as well as an increase in trajectory length due to higher swimming speeds.  For $k_{sp} = 0.01$ and $k_{sp} = 0.2$, we observe gradual changes in the swimming direction, while for $k_{sp} = 2$ the trajectories contain many loops and sudden turns.  For $k_{sp}=8$, the swimmer also changes direction quite often, however, it often becomes trapped by the obstacles long before it reaches 200 periods of undulation, resulting in short, terminated trajectories.  

We observe similar, but less dramatic changes in the trajectories when $k_{sp}$ is fixed and $\varphi$ is increased.  Fig. \ref{fig:TrajectoriesAf} shows trajectories for $k_{sp} = 0.2$ and $\varphi = 0.15$, $0.25$, $0.35$ and $0.45$.  As $\varphi$ increases, we see that the lengths of the 200$T$ trajectories increase, as does the tendency for the swimmer to change direction.  We, however, do not see the very tortuous trajectories observed at the highest values of $k_{sp}$, nor do we observe the swimmer becoming trapped, even at high densities.  

\subsection{Stochastic model}

From the long time simulations presented above, we saw how tether stiffness and obstacle density affected the trajectories exhibited by the swimmer due to changes in swimming speed and induced velocity fluctuations.  In order to better quantify long-time swimmer behaviour and how it varies with environmental parameters, we employ a stochastic model that uses as input data from short-time simulations.  In this model, the swimmer centre-of-mass position, $\bm{X} = (X, Y)$, and orientation, $\hat{\bm{p}} = (\cos \theta, \sin \theta)$, are described by the stochastic differential equation
\begin{align}
d \begin{bmatrix}
X \\ Y \\ \theta 
\end{bmatrix} = \langle V_{p} \rangle \begin{bmatrix}
\cos \theta \\ \sin \theta \\ 0 
\end{bmatrix} dt + \sqrt{2 \tau} \bm{R}(\theta) \bm{C}^{1/2} d \bm{B}, \label{eq:StochEquationSolve}
\end{align}
where 
\begin{align}
\bm{R}(\theta) =
  \begin{bmatrix}
    \cos \theta & -\sin \theta & 0  \\ 
    \sin \theta & \cos \theta & 0 \\
    0 & 0 & 1
  \end{bmatrix},
\end{align}
is the rotation matrix from the body to lab frames, $\bm{C}^{1/2}$ is the the Cholesky factorization of the covariance matrix
\begin{align}
\bm{C} =
  \begin{bmatrix}
   C_{pp} & 0 & C_{p\Omega}  \\ 
    0 & C_{nn} & C_{n\Omega}  \\
     C_{p\Omega} & C_{n\Omega} & C_{\Omega\Omega}
  \end{bmatrix},
\end{align}
and $d\bm{B}$ is the increment of a vector of independent Wiener processes.  The parameter $\tau$ describes the short correlation time of the velocity fluctuations due to collisions with the obstacles.  The effects of the obstacles on swimmer motion are incorporated into the model by using the values of $\langle V_{p} \rangle$ and $\bm{C}$ computed from the full simulations.  We note that the stochastic model resembles that used to describe active Brownian particles (ABPs) \cite{Volpe2013,Volpe2014} and bacteria subject to rotational diffusion \cite{Lauga2011}, however, here, the covariance matrix is both anisotropic ($C_{pp} \neq C_{nn}$) and the random velocities and angular velocities are correlated through the non-zero entries $C_{n\Omega}$ and $C_{p\Omega}$.  Additionally, unlike ABPs and bacteria where the noise term can be attributed to thermal fluctuations, or inherently random fluctuations in the bacteria's flagellar movements, in our case, the fluctuations are due to collisions with the obstacles and, as a result, are accompanied by changes in the swimming speed. 

\subsection{Autocorrelation functions and means-squared displacement}

\begin{figure}[h]
\centering
 \includegraphics[width=7in]{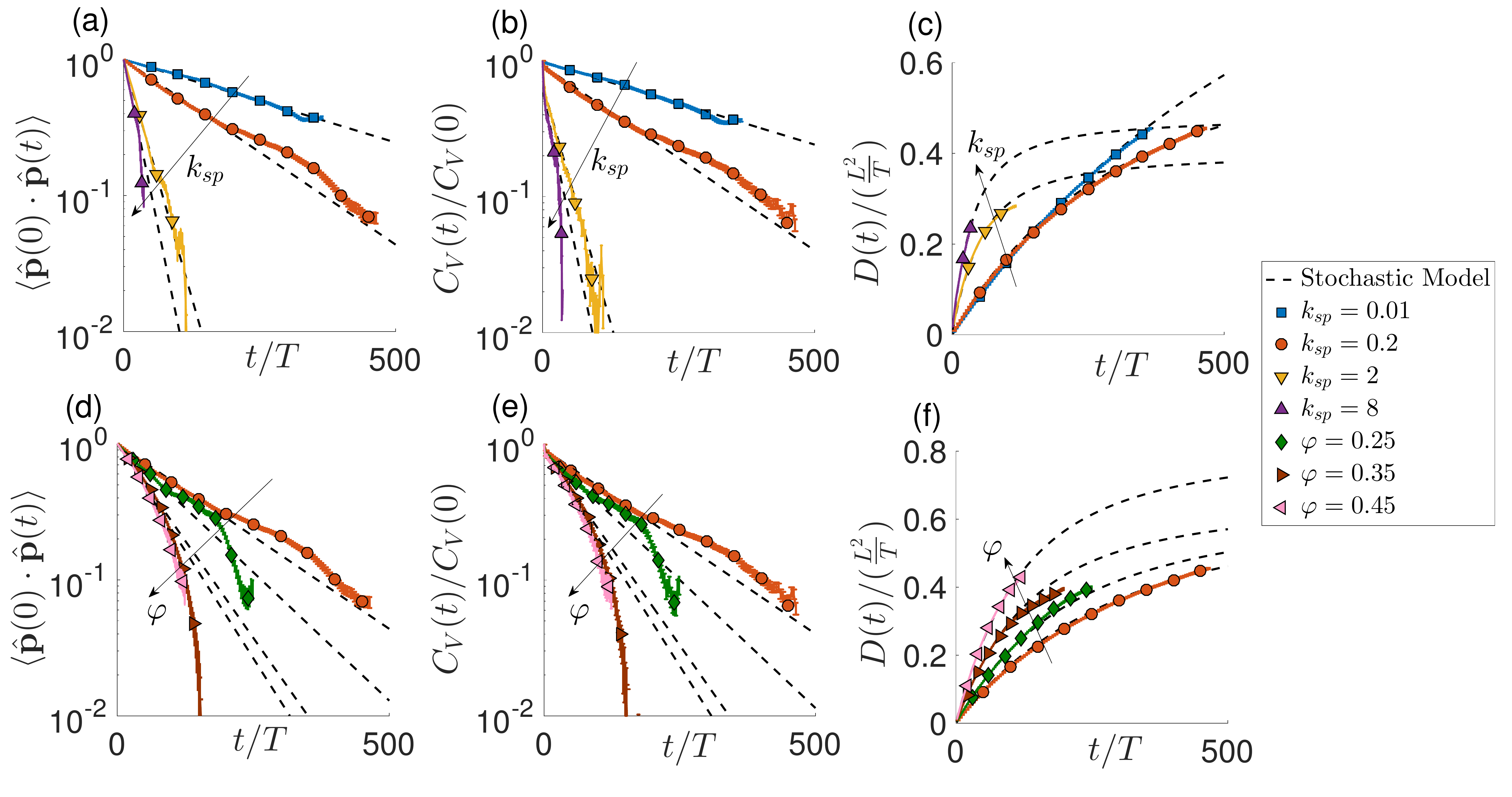}
\captionsetup{width=0.9\textwidth}
\caption{\sf Autocorrelation functions and time-dependent diffusion coefficient from long-time simulations (solid lines) and the stochastic model (dashed) for $\varphi = 0.15$ and $k_{sp} = 0.01, 0.2, 2$ $\&$ $8$. The panels show the (a) Orientation autocorrelation function, $\langle {\hat {\bm p}} (0) \cdot {\hat {\bm p}} (t) \rangle $, (b) Velocity autocorrelation function, $C_{V}(t)$, and (c) the time-dependent diffusion coefficient, $D(t)$.  Panels (d) -- (f) show these same quantities from long-time simulations and the stochastic model for $k_{sp} = 0.2$ and $\varphi = 0.15, 0.25, 0.35$ $\&$ $0.45$.}
\label{fig:ACFs}
\end{figure}

From the stochastic model, we can obtain expressions for the swimmer orientation and velocity autocorrelation functions, as well as the mean squared displacement.  We can then relate these quantities back to the environmental parameters $k_{sp}$ and $\varphi$ to assess how they affect swimmer motion at longer times.  The orientation autocorrelation function (OACF) can be found by integrating the equation for $\theta$ to give
\begin{equation}
\langle {\hat {\bm p}} (0) \cdot {\hat {\bm p}} (t) \rangle = e^{-C_{\Omega\Omega}\tau t}, \label{eq:OrientationACF}. 
\end{equation}
The details of this calculation are presented in the Supplementary Material.  We see that the OACF decays exponentially with a correlation time given by $\tau_c = (C_{\Omega\Omega}\tau)^{-1}$.  We also observe that even though the matrix $\bm{C}$ contains off-diagonal entries, only the diagonal entry $C_{\Omega\Omega}$ affects the OACF.  

In Fig. \ref{fig:ACFs}a we compare the OACF given by the stochastic model with that computed from full simulations with $\varphi = 0.15$ and $k_{sp} = 0.01, 0.2, 2.0$ and $8$.  In each case, the simulation data is well described by the exponential OACF given by Eq. (\ref{eq:OrientationACF}) with the correlation times decreasing with $k_{sp}$, going from $\tau_{c} = 356.0T$ for $k_{sp} = 0.01$ down to $\tau_{c} = 22.3T$ for $k_{sp} = 8$.  This is consistent with our observations of the trajectories where the swimming direction changes more often and more drastically at higher tether stiffnesses.  Using the values of $\tau_c$ and $C_{\Omega\Omega}$, we can obtain $\tau$, the collision correlation time.  For these four cases, we have $\tau = 0.93 T$ ($k_{sp} = 0.01$), $\tau = 0.49 T$ ($k_{sp} = 0.2$), $\tau = 0.57 T$ ($k_{sp} = 2$), and $\tau = 0.53 T$ ($k_{sp} = 8$), indicating that the correlation time associated with swimmer-obstacle collisions is on the order of a single period of undulation.  These values of $\tau$ are used for subsequent comparisons presented below.  

Along with the OACF, from the stochastic model we can also compute the velocity autocorrelation function (VACF) 
\begin{align}
C_{V}(t) = \langle \bm{V}(t) \cdot \bm{V}(0) \rangle =& \left(\langle V_{p} \rangle^{2} + 2 \langle V_{p} \rangle\tau C_{n\Omega} \right) e^{-C_{\Omega\Omega}\tau t}  + 2\tau (C_{pp} + C_{nn}) \delta(t) \nonumber\\
&+ \tau^{2} (C_{p\Omega}^{2} +  C_{n\Omega}^{2}) \mathds{1}_{ \{ t = 0 \}},
\label{eq:VACF} 
\end{align}
where, formally, $\bm{V} = d\bm{X}/dt$, $\delta(t)$ is the Dirac delta function, and $\mathds{1}_{ \{ t = 0 \} }$ is the function that is $1$ at $t=0$ and $0$ for $t >0 $.  The details of this calculation may also be found in the Supplementary Materials.  The VACF from the stochastic model and long-time simulations are shown in Fig. \ref{fig:ACFs}b for $\varphi = 0.15$.  As with the OACF, the stochastic model predicts that the VACF decays exponentially with correlation time $\tau_c = (C_{\Omega\Omega}\tau)^{-1}$ and reproduces the VACF determined from the long-time simulations for each value of $k_{sp}$.  Along with the exponential decay, we observe a sharp initial drop in the VACF corresponding to the additional short-time correlations appearing in Eq. (\ref{eq:VACF}). 

Finally, from the stochastic model, we compute the swimmer's time-dependent diffusion coefficient, $D(t) = \langle ({\bf X}(t) - {\bf X}(0))^{2} \rangle/4t$, 
\begin{equation}
\begin{split} 
D(t) = \frac{\langle V_{p} \rangle^{2}}{2C_{\Omega\Omega}\tau} \left( 1 - \frac{1}{C_{\Omega\Omega} \tau t}(1 - e^{-C_{\Omega\Omega}\tau t}) \right) + \frac{\tau}{2}(C_{pp} + C_{nn}) \\ +  \frac{\langle V_{p} \rangle C_{n\Omega}}{C_{\Omega\Omega}} \left( 1 - \frac{1}{C_{\Omega\Omega}\tau t}(1 - e^{-C_{\Omega\Omega}\tau t}) \right).
\end{split} \label{eq:MSDExpression}
\end{equation}
Again, the details of the calculation can be found in the Supplementary Material.  Fig. \ref{fig:ACFs}c shows $D(t)$ for $\varphi = 0.15$ given by both the long-time simulations and Eq. (\ref{eq:MSDExpression}), and we again see close agreement between the simulations and stochastic model.  As the swimming speed increases with $k_{sp}$, we observe a more rapid initial growth of $D(t)$ in stiffer environments.  For higher values of $k_{sp}$, we see the onset of diffusive behaviour as $D(t)$ approaches a constant value at $t \approx 300T$.  For lower $k_{sp}$, $D(t)$ grows linearly and, due to the very long correlation times found for these environments, even after 500 undulation periods it has yet to even begin leveling off.  

Figs. \ref{fig:ACFs}d-f, show the OACF, VACF, and $D(t)$, from the stochastic model and full simulations for $k_{sp} = 0.2$ and $\varphi = 0.15$, $0.25$, $0.35$, and $0.45$.  Generally speaking, we find agreement between the stochastic model and the full simulations for these environmental parameters, though for higher obstacle densities, we do see some discrepancy at large times, where the correlations are found to decay rapidly and faster than the exponential predicted by the model.  From the OACF, we observe that as $\varphi$ increases, the correlation time decreases from $\tau_c = 159.6T$ at $\varphi = 0.15$ to $\tau_c = 68.9T$ when $\varphi = 0.45$.  As a result, VACF decays more rapidly as $\varphi$ increases and $D(t)$ reaches constant values sooner.  We note that these changes are not as dramatic as those seen when increasing $k_{sp}$, and the notable initial drop in the VACF due to the $\delta$-function is essentially absent in these cases.  

\subsection{Effective diffusion coefficient and correlation times}
\begin{figure}[h]
\centering
\includegraphics[width=5in]{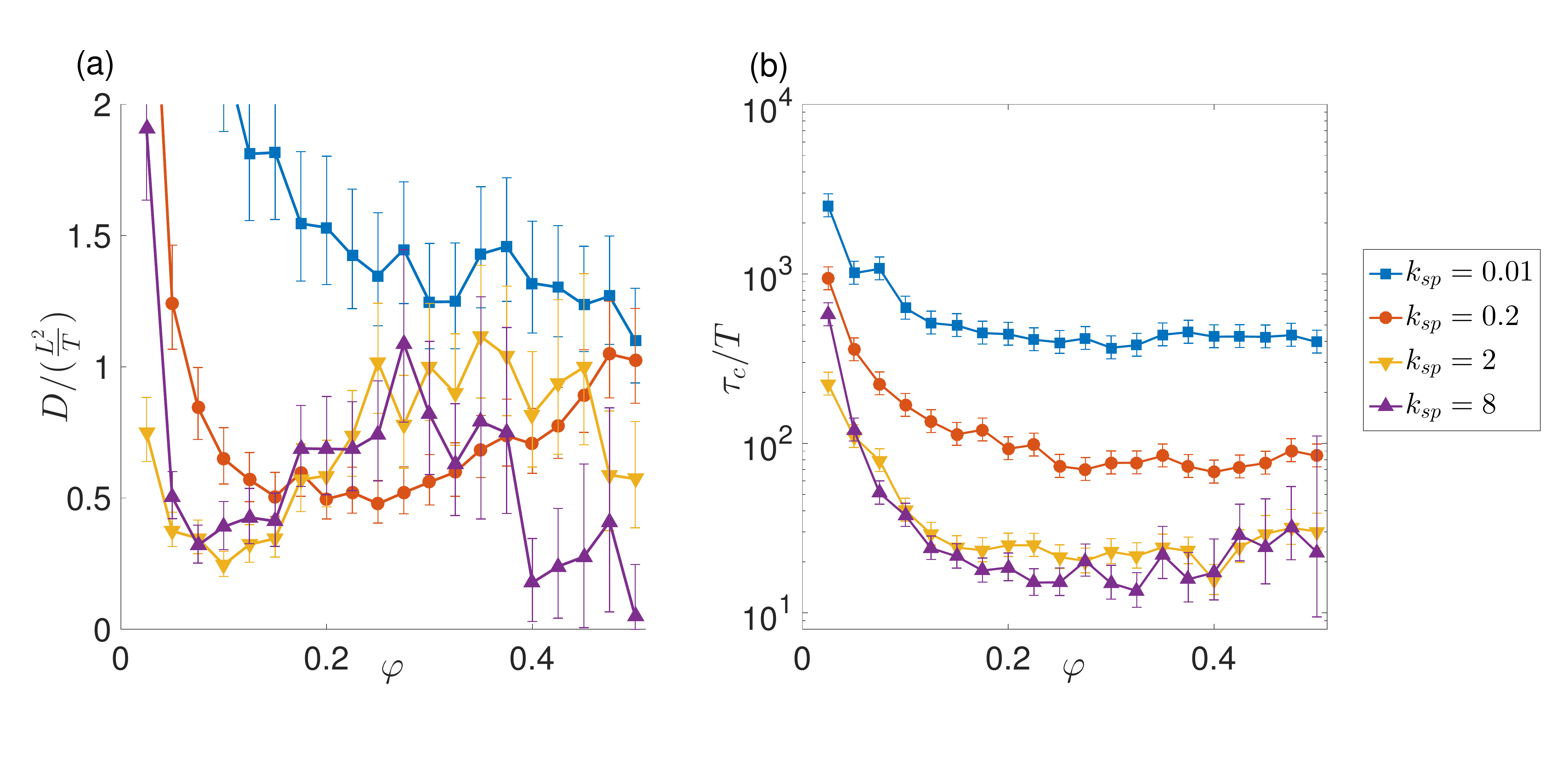}
\captionsetup{width=0.9\textwidth}
\caption{\sf (a) Effective diffusion coefficient, $D$, and (b) correlation time, $\tau_{c}$, given by the stochastic model for $k_{sp} = 0.01, 0.2, 2$ $\&$ $8$.  The error bars indicate the 95\% confidence intervals computed using a local sensitivity analysis \cite{ku1966notes} based on the expressions for $D$ and $\tau_c$ and the confidence intervals for the values of $\langle V_{p}\rangle$ and entries of $\bm{C}$.}
\label{fig:DandTauc}
\end{figure}

From the stochastic model, we can obtain the effective swimmer diffusion coefficient 
\begin{align}
D = \lim_{t \rightarrow \infty} D(t) = \frac{\langle V_{p} \rangle^2}{2\tau C_{\Omega\Omega}} +  \frac{\tau}{2} (C_{pp} + C_{nn}) + \frac{\langle V_{p} \rangle C_{n\Omega}}{C_{\Omega\Omega}}, \label{eq:effD}
\end{align}
which characterises the diffusive motion of a single swimmer at long-times, but also provides a measure of how a dilute population of swimmers would spread with time.  We see that the effective diffusion consists of three terms that depend on $\langle V_{p} \rangle$ and the entries of $\bm{C}$.  The first term is the contribution that results from the coupling of rotational diffusion induced by collisions and forward locomotion \cite{Lauga2011,Volpe2014,Zeitz2017}.  The second term arises due to the velocity fluctuations induced by collisions with the obstacles, while the third term is an additional contribution due to the covariance between translational and rotational motion as a consequence of $C_{n\Omega} \neq 0$.     

Fig. \ref{fig:DandTauc}a shows the effective diffusion coefficient as a function of $\varphi$ for $k_{sp} = 0.01, 0.02, 2$ and $8$.  Here, the values for $\langle V_{p} \rangle$ and $\bm{C}$ are taken from the short-time simulations presented in Section \ref{sec:motion} with the trapped periods removed from the averaging (see Supplementary Materials).  We also have assumed that $\tau = 0.6T$ for all cases.  For purposes of discussion, the correlation times, $\tau_c = (C_{\Omega\Omega}\tau)^{-1}$, corresponding to each case are shown in Fig. \ref{fig:DandTauc}b.  We examine the contribution of each of the three terms appearing in Eq. (\ref{eq:effD}) and find that the overwhelming contribution to $D$ for each case is the term $\langle V_{p} \rangle^2/(2C_{\Omega\Omega}\tau)$.  It's lowest contribution is found for $k_{sp} = 8$ and $\varphi = 0.5$, where it still accounts for $93.7$\% of $D$.  Thus, the effective swimmer diffusion is due primarily to a coupling between swimming and rotational diffusion.

At low obstacle densities, or low tether stiffnesses, we find that $D$ can be quite large values due to the lack fluctuations and long correlation times found for these environments.  For moderate densities where fluctuations are more significant, we find that, that the value of $D$ appears to be independent of the tether stiffness.  Thus, the increases in swimming speed that occur when $k_{sp}$ increases are balanced the accompanying increases in rotational diffusion as to keep $\langle V_{p} \rangle^2/(2C_{\Omega\Omega}\tau)$ constant.  We note, however, that the increase in rotational diffusion does lead to large differences in correlation times with $\tau_c \approx 100T$ for $k_{sp} = 0.2$ and $\tau_c \approx 20T$ for $k_{sp} = 2$ and $8$.  Thus, even though the diffusion coefficient may be the same, it will take longer for the swimmer to exhibit diffusive behaviour in more compliant environments.  

\section{Swimmers are trapped by stiff, dense environments}\label{sec:trapping}

\begin{figure}[h]
\centering
 \includegraphics[width=5in]{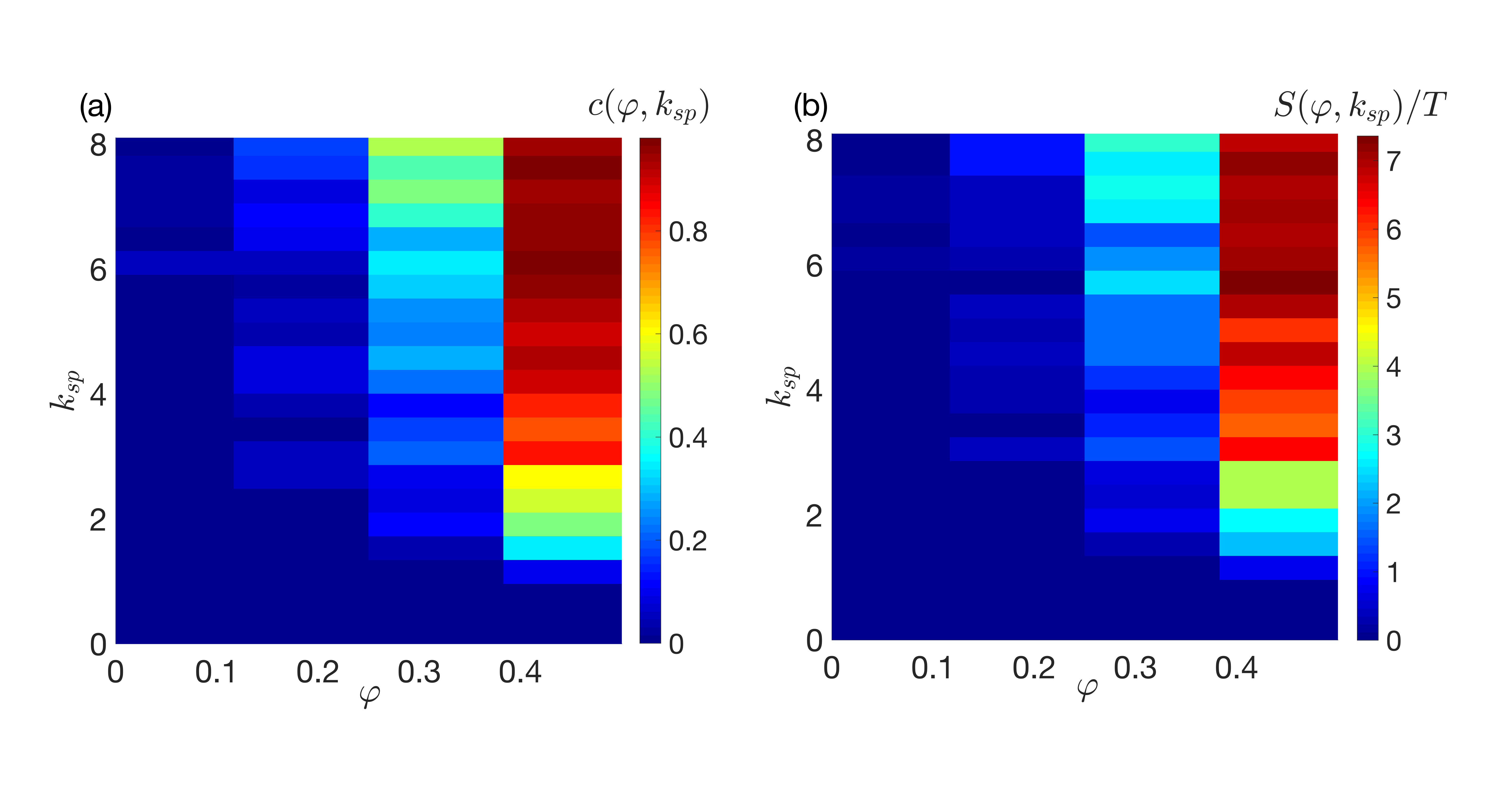} % EPS plot
 \captionsetup{width=0.9\textwidth}
\caption{\sf (a) Fraction of trapped swimmers $c(\varphi, k_{sp})$ and (b) average time trapped $S(\varphi, k_{sp}) / T$ from short-time simulations over the $\varphi-k_{sp}$ parameter space.  At high $\varphi$ and $k_{sp}$, we see that the swimmer is both more likely to be trapped and be trapped for longer periods of time.}
\label{fig:Trapping}
\end{figure}

While we can characterise the diffusion coefficient using our short time data and the stochastic model, it is important to recall that at high tether stiffness and obstacle density the swimmer becomes trapped by the environment, perhaps even before the onset of diffusive behaviour.  In our simulations, when trapping occurs (see video provided as electronic Supplementary Material), we find that though the swimmer continues to undulate, it collides with the same set of obstacles without moving forward.  Since our simulations are deterministic, once the swimmer is trapped, it remains trapped indefinitely.  Trapping in our simulations is linked to obstacle interactions that modify the swimmer's waveform and prevent it from making any forward motion.  This is in contrast with previous studies where trapping is instead linked to the swimmer moving in closed, periodic trajectories around a particular set of obstacles \cite{Majmudar2012,Munch2016,Chamolly2017,Takagi2014}.

To quantify the likelihood of trapping, we compute for different values of $k_{sp}$ and $\varphi$ the trapping fraction, $c(\varphi, k_{sp}) = N_{trap}(\varphi, k_{sp})/ N_{sim}$, where $N_{trap}(\varphi, k_{sp})$ is the number of simulations in which the swimmer becomes trapped before $10T$ and $N_{sim}$ is the number of simulations that are run for each case.  For most cases, $N_{sim} = 50$, however, for the cases where $k_{sp} = 2$ and $k_{sp} = 8$, we have $N_{sim} = 90$ as we also use our simulation results from Section \ref{sec:motion} to compute $c(\varphi, k_{sp})$.  Fig. \ref{fig:Trapping}a shows the trapping fraction over the $\varphi$-$k_{sp}$ parameter space.  We find that trapping only occurs when $k_{sp} \gtrsim 1$, or rather, when the tether stiffness is greater than the stiffness of the swimmer.  At high obstacle densities where $\varphi = 0.45$, we see a very sharp transition at $k_{sp} \approx 1$ where the trapping fraction changes rapidly from $c =0$ to $c \approx 1$.  For fixed $k_{sp}$, the trapping fraction increases with $\varphi$ provided $k_{sp} \gtrsim 1$.  For the highest tether stiffnesses, we found that the swimmer can become trapped at obstacle densities as low as $\varphi = 0.1$.  Additionally, trapping can occur at times greater than $10T$.  In fact, from our long-time simulations with $\varphi = 0.15$ and $k_{sp} = 8$, we found that all swimmers would eventually become trapped by the environment.   

Not only is the swimmer more likely to be trapped in stiffer, denser environments, but it is also more likely to be trapped sooner.  Fig. \ref{fig:Trapping}b shows the average time trapped, $S(\varphi, k_{sp}) = (1/N_{sim})\sum_{n=1}^{N_{sim}}T_{trap}^{n}(\varphi, k_{sp})$, where $T_{trap}^{n}(\varphi, k_{sp})$ is the time the swimmer in simulation $n$ is trapped during the last $8T$ of the simulation.  We see that $S(\varphi, k_{sp})$ follows the same trends in both $\varphi$ and $k_{sp}$ as $c(\varphi, k_{sp})$, with the longest times trapped occurring at the largest values of $\varphi$ and $k_{sp}$.

\begin{figure}[h]
\centering
 \includegraphics[width=5in]{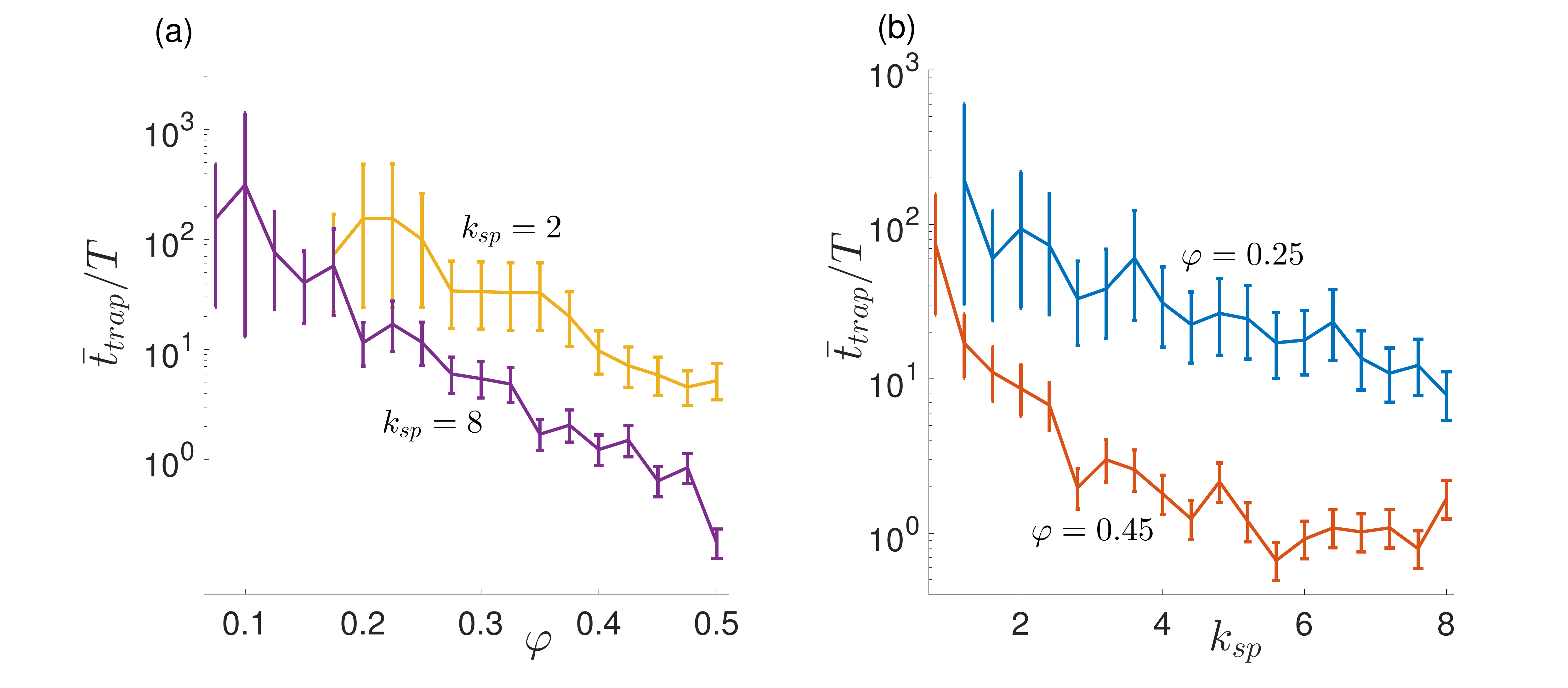} % EPS plot
 \captionsetup{width=0.9\textwidth}
\caption{\sf The maximum likelihood estimate of the mean trapping time, $\bar{t}_{trap}$, as a function of (a) $\varphi$ for $k_{sp} = 2$ and $k_{sp} = 8$, and (b) $k_{sp}$ for $\varphi = 0.25$ and $\varphi = 0.45$.}
\label{fig:TrappingTimes}
\end{figure}

To further quantify trapping, we compute using the short-time data presented in Section \ref{sec:motion} the maximum likelihood estimate \cite{Lawless2011},
\begin{equation}
\bar{t}_{trap} = \frac{1}{N_{trap}} \sum_{n=1}^{N_{sim}} t^\star_{n} , \label{eq:MLE}
\end{equation}
where $N_{sim}$ is again the total number of simulations, $N_{trap}$ in the number of simulations where trapping occurs before $8T$, and $t^\star_{n} = \min(t_{trap,n}, 8T)$ with $t_{trap,n}$ being the time the swimmer in simulation $n$ is trapped.  In Eq. (\ref{eq:MLE}), it is assumed that for each $k_{sp}$ and $\varphi$ the trapping times are distributed exponentially with trapping rate $\lambda = 1/\bar{t}_{trap}$.  This assumption is checked for consistency by comparing the average mean squared distance travelled from full simulations with those obtained using the stochastic model and an exponential distribution of trapping times (see Supplementary Material).     Fig. \ref{fig:TrappingTimes}a shows $\bar{t}_{trap}$, as a function of $\varphi$ for fixed values of $k_{sp}$.  With $k_{sp}$ fixed, we find that the average trapping time decays exponentially with obstacle density.  Fitting the data with a function of the form $c_0\exp(-c_1 \varphi)$ yields $c_1 = 11.32$ for $k_{sp} = 2$ and $c_1 = 15.0$ for $k_{sp} =8$, indicating that the decay rate does not depend strongly on tether stiffness.  For $k_{sp} = 8$, the average trapping times decrease from approximately 100 undulations periods at lowest area fractions to a just single period at $\varphi = 0.45$.  In addition, we see that for these cases, the average trapping times are comparable to the correlation time $\tau_c$ from Fig. \ref{fig:DandTauc}.  As a result, it is likely that swimmers moving through these environments would be trapped before their spreading is described by diffusion alone.  From Fig. \ref{fig:TrappingTimes}b, we see also that the trapping time decreases with tether stiffness when the obstacle density is fixed.  For $\varphi = 0.25$, there is a gradual exponential decay in the trapping time, while for $\varphi = 0.45$, the decay is more rapid going from $100T$ at $k_{sp} \approx 1$ to just a single period at $k_{sp} \approx 4$, indicating that in dense environments, swimmers will often be trapped instantaneously.    

\section{Discussion and conclusions}

In this paper, we presented results from a series of simulations of an undulatory swimmer moving through an environment consisting of fluid and a 2D arrangement of rigid spherical obstacles that are connected by linear springs to random points in space.  Our results demonstrate how the discrete interactions between a swimming body and other microscopic structures, such as polymers or filaments, immersed in the surrounding fluid affect swimmer motion.  In particular, we show not only how the presence of the obstacles can often enhance the swimming speed, but also how the discrete interactions lead to fluctuations in the swimmer's translational and angular velocities.  These fluctuations, coupled with the swimming velocity, lead to diffusive behaviour at long times, which we can quantify using a stochastic model.  We also show how obstacles can hinder motion, leading to swimmer trapping, particularly in dense environments with stiffnesses greater than that of the swimmer.  Increasing obstacle density provides a simple mechanism for increasing trapping of swimmers and is consistent with the observation of density variations of cervical mucus over the female cycle \cite{Suarez2006}.  Additionally, our results indicate that a minimum stiffness of the environment is also required, and only then can variations with density occur.

While we have studied here how phenomena such as trapping, enhanced locomotion, and effective diffusion vary with environmental properties, it is also of interest to investigate further how these phenomena change with swimmer's gait, or propulsion strategy.  It has been proposed \cite{Suarez2006,Holt2009,Holt2015} that sperm selection based on gait by cervical mucus may play a role in allowing only the most genetically viable sperm to reach the egg.  Additionally, in continuum descriptions of viscoelastic fluids, it is known that rear versus front actuation by undulatory swimmers leads to greater enhancement of swimming speed \cite{Thomases2014}.  In our simulations, the swimmer's front-actuated gait is fixed and based on that of {\it C. elegans}.  Understanding if and how our results carry over to swimmers with different waveforms, including helical ones \cite{Zhang2018,Zoettl2017}, could give some indication of how the fluid microstructure can segregate populations of different swimmers based on how they move.  In fully 3D arrangements of filaments, filament alignment and anisotropy may play a role, potentially even to guide the swimming cells in a particular direction \cite{Chretien2003}, while in filament networks, connectivity and cross-linking could lead to increased trapping.

Additionally, interactions between swimmers are modified as a result of the immersed microstructure.  It has been observed \cite{Tung2017} that the inclusion of viscoelasticity leads to the formation of coherent groups of moving sperm cells.  In heterogeneous environments, the complexity of the interactions with the immersed microstructure can introduce further effects, such as the local rearrangement of obstacles, hydrodynamic screening of induced flows by the microstructure, or perhaps long distance propagation of elastic deformations through obstacle collisions.  These effects, as well as their coupling with biologically relevant phenomena, such as chemotaxis, can provide further changes in swimmer behaviour as a result of their direct interactions with immersed structures.  

\section{Acknowledgements}
The authors would like to thank Professor Michael Shelley and Professor Pierre Degond for many useful discussions.  We also thank Noah Brenowitz, Anton Glazkov, and Samuel Colvin for preliminary work during undergraduate projects.  EEK gratefully acknowledges support from EPSRC grant EP/P013651/1.

\bibliographystyle{unsrt} 
\bibliography{References}  

\end{document}